\documentstyle[12pt,preprint,aps,epsf]{revtex}
\tightenlines
\newcommand{\comment}[1]{}
\begin{document}
\draft

\title{
Numerical Evidence for Divergent Burnett Coefficients
}

\author{Russell K. Standish}
\address{
High Performance Computing Support Unit\\
The University of New South Wales
}

\maketitle

\begin{abstract}
  In previous papers [Phys. Rev. A {\bf 41}, 4501 (1990), Phys. Rev. E
  {\bf 18}, 3178 (1993)], simple equilibrium expressions were obtained
  for nonlinear Burnett coefficients. A preliminary calculation of a
  32 particle Lennard-Jones fluid was presented in the previous paper.
  Now, sufficient resources have become available to address the
  question of whether nonlinear Burnett coefficients are finite for
  soft spheres. The hard sphere case is known to have infinite
  nonlinear Burnett coefficients (ie a nonanalytic constitutive
  relation) from mode coupling theory. This paper reports a molecular
  dynamics caclulation of the third order nonlinear Burnett
  coefficient of a Lennard-Jones fluid undergoing colour flow, which
  indicates that this term is diverges in the thermodynamic limit.
\end{abstract}
\narrowtext

\section{Introduction}

Ever since the Green-Kubo formalism for calculating the linear
transport coefficients was developed, there has been interest in a
corresponding theory for the nonlinear Burnett coefficients. The
discovery of long-time tails in the velocity autocorrelation function
by Alder and Wainwright \cite{Alder-Wainwright70} indicated that the
hydrodynamic transport coefficients do not exist in two dimensions,
but do exist in three dimensions. By applying mode-coupling theories,
Ernst {\em et al.} \cite{Ernst-etal78} showed that the relation
between stress and strain rate should be $P_{xy}\propto|\gamma
|\ln|\gamma |$ for hard disks and $P_{xy}=-\eta \gamma +c|\gamma
|^\frac {3} {2} $ for hard spheres, which are non-analytic
constitutive relations.  Similar results were obtained by Kawasaki and
Gunton\cite{Kawasaki-Gunton73} for incompressible fluids (which is a
particular case of a hard sphere fluid), although criticised later by
Brey et. al.\cite{Brey-etal81}. It should be pointed out that the
linear Burnett coefficients are known to be
divergent\cite{Ernst-etal75,Wood75}, and in light of the
linear coefficient results, it is generally assumed that the nonlinear
coefficients should be divergent as well for soft particle systems.
Brey et. al.\cite{Brey-etal81} claim to show the divergence of
nonlinear coefficients in a followup paper (ref. 11 in that paper),
yet this paper never appeared in the literature. Therefore, there is
considerable interest in a molecular dynamics simulation of a soft
particle system to see if the hard sphere results generalise.

In a paper by Evans and Lynden-Bell \cite{Evans-Lynden-Bell88},
equilibrium fluctuation expressions for inverse Burnett coefficients
were derived for the colour conductivity problem.  The coefficients,
$B_i$, give a Taylor series representation of a nonlinear transport
coefficient $L$, in terms of the thermodynamic force $F$.  Thus if a
thermodynamic flux $J$ is written in terms of the coefficient's
defining constitutive relation as $\langle J\rangle=L(F)F$, then the
Burnett coefficients are related by $L(F)=B_0+B_1F+B_2F^2+\cdots$.  In
order to derive closed form expressions for the Burnett coefficients,
it was found necessary to work in the Norton ensemble, in which the
flux $J$, rather than the thermodynamic force $F$ was the independent
variable.  The constitutive relation in this case is $\langle
F\rangle= {\cal L}(J)J= {\cal B}_0+ {\cal B}_1J+\cdots$.  In the
thermodynamic limit, we may write $ {\cal L}(J)= L^{-1}(J)$, and so
the non-linear Burnett coefficients can be computed by inverting the
series.

Evans and Lynden-Bell \cite{Evans-Lynden-Bell88} applied constant
current dynamics to a canonical ensemble with the currents distributed
about an average current $J_0$.  This allowed the derivation of a
transient time correlation function for the non-equilibrium phase
average $\langle F\rangle$.  It was then a simple matter to compute
the derivatives of $\langle F\rangle$ with respect to the average
current $J_0$, as the constant current propagator commutes with the
derivative operator.  However, this method appeared to be limited to
colour currents, for which an appropriate canonical distribution could
be found.  In a previous paper \cite{Standish-Evans90} we show that
this method can be applied to the situation of an arbitrary
thermodynamic flux.  Later, \cite{Standish-Evans93} we showed that
this transient time correlation expression can be expressed in terms
of an average over an equilibrium simulation, reducing the calculation
required by two orders of magnitude. At the time, computational
resources were not sufficient to establish whether this expression is
finite in the limit as $t\rightarrow\infty$, or in the thermodynamic
limit. In this paper, we present computational results of colour
conductivity in a Lennard-Jones system, harnessing 4 supercomputers
simultaneously over a period of 18 months, that show distinct evidence
that ${\cal B}_2=\infty$.

In order to avoid confusion, it should be noted that the term ``colour
diffusion'' is sometimes used in the sense of the diffusion of colour
labels attached to otherwise colour blind particles in the complete
absence of applied external fields \cite{Holian-Wood73}. In this
approach if the colour label attached to a particle is ignored, the
system remains at equilibrium. This is manifestly a linear process.
In the model we consider all the particles interact with an external
colour sensitive external field and this allows the possibility of a
nonlinear response. It might also be pointed out the the colour field
we consider here is independent of both position and time so that the
{\em linear} Burnett coefficients do not play a role.

\section{The Simulation}

The simulation was performed using the colour 
conductivity model described in Evans and 
Lynden-Bell \cite{Evans-Lynden-Bell88}. The intermolecular 
potential was taken to be the Lennard-Jones potential, which has an attractive 
component due to van der Waals interaction, and a repulsive hard core that 
goes as $r^{-12}$:
\begin{displaymath}
V(r)=4\varepsilon \left(
  \left(
    \frac
      {\sigma }
      {r}
    \right)^{12}-
  \left(
    \frac
      {\sigma }
      {r}
    \right)^6
  \right).
\end{displaymath}
In what follows, every quantity will be given in reduced units, in which 
$\varepsilon =\sigma =m=1$. This model has been well
studied, and can be related physically to some noble gases like argon.

The system was simulated at 3 different system sizes (32, 108 and 256
particles) using a periodic boundary condition to minimise boundary
effects. The state point chosen had a temperature of 1.08 and density
of 0.85. Considerable information was already known about this system
at that state point \cite{Evans-Morriss85}.

The equations of motion are just that of the Nos\'e-Hoover thermostat,
with an additional flux statting term. This generates a canonical
ensemble:
\begin{eqnarray} \label{eq. motion}
\dot{\bf q}_i&=&\frac{{\bf p}_i}{m},\nonumber\\*
\dot{\bf p}_i&=&{\bf F}_i+e_i\hat{\bf x}\lambda-\alpha{\bf p}_i,\nonumber\\*
\dot{\alpha}&=&\frac{3Nk_B}{Q_\alpha}(T-T_0),\nonumber\\*
\dot{\lambda}&=&\frac{N}{Q_\lambda}(J-J(t=0)),
\end{eqnarray}
where ${\bf F}_i$ are the intermolecular forces, $e_i=\pm1$ are the
colour charges, $T=\sum\frac{mp_i^2}{3Nk_B}$, $T_0=\langle T\rangle$ and
$J=\sum_i\frac{p_{xi}e_i}{Nm}$ is the colour current.

The feedback parameter $Q_\lambda$ was chosen equal to 4.74 for the
108, 256 and one of the 32 particle runs. Because $Q_\lambda$ should
be an extensive quantity, the 32 particle run was repeated at
$Q_\lambda=32\times4.74/108=1.4$. The Nos\'e-Hoover thermostat
parameter $Q_\alpha$ was chosen to be $0.31N$.  The values of these
parameters were chosen to give optimal convergence of the linear
response function. There is no real reason for them to be optimal for
non-linear response functions.

When the flux is fixed in this manner, the ensemble is termed a Norton
ensemble. When the thermodynamic force is fixed, then it is termed a
Th\'evenin ensemble by analogy with electrical circuits\cite{Evans-Morriss85}.
We have recently given
a statistical mechanical proof of the macroscopic equivalence of the Norton
and Th\'evenin representations of a nonequilibrium system \cite{Evans93}.

Recall that Transient Time Correlation Functions for evaluating the
inverse cubic Burnett coefficient ${\cal B}_2$ is given in
Ref. \cite{Evans-Lynden-Bell88}: 
\widetext
\begin{eqnarray} \label{B2}
{\cal B}_2=
\frac
  {3N\beta }
  {\langle\Delta J^2\rangle^2}
\int_{0}^{\infty}\langle \lambda(s)\lambda(0)(\Delta J^2-\langle\Delta
J^2\rangle)\rangle ds. 
\end{eqnarray}
\narrowtext where $\lambda(s)$ is the additional phase variable
(defined in eq. (\ref{eq. motion}) corresponding to a colour force of
a system at time $s$ along a trajectory and $J$ is the colour current
at the origin of that trajectory. As the system is at equilibrium (in
the canonical ensemble), after a correlation time has passed, the
system's configuration is effectively randomised, and may be used as a
new trajectory origin.
The correlations between different successive states of the
equilibrium simulation can be easily seen by examining something like
the velocity autocorrelation function (see Fig 7.1 of
\cite{Hansen-McDonald86} for examples). The correlation time for this
system is about 1.

\section{Results}

Because the relevant quantity is an ensemble average, a very effective
parallelisation strategy is to run a separate copy of the system on
each processor, compute the TTCF on each processor, then average over
the entire set of processors, weighting for the number of timesteps
executed on each processor. Further computational details of this
experiment have been reported in \cite{Standish97c}. Whilst the
results of this experiment would appear meagre compared with the
computational resources used to compute it, it should be pointed out
that this computation was conducted at the lowest priority on these
machines, using idle CPU cycles. 

Having a set of approximations also allows one to calculate the standard
error of the TTCF. These are shown as error bars in Figures
\ref{Fig1}--\ref{Fig6}

%

\begin{figure}
\begin{center}
\mbox{}\epsfbox{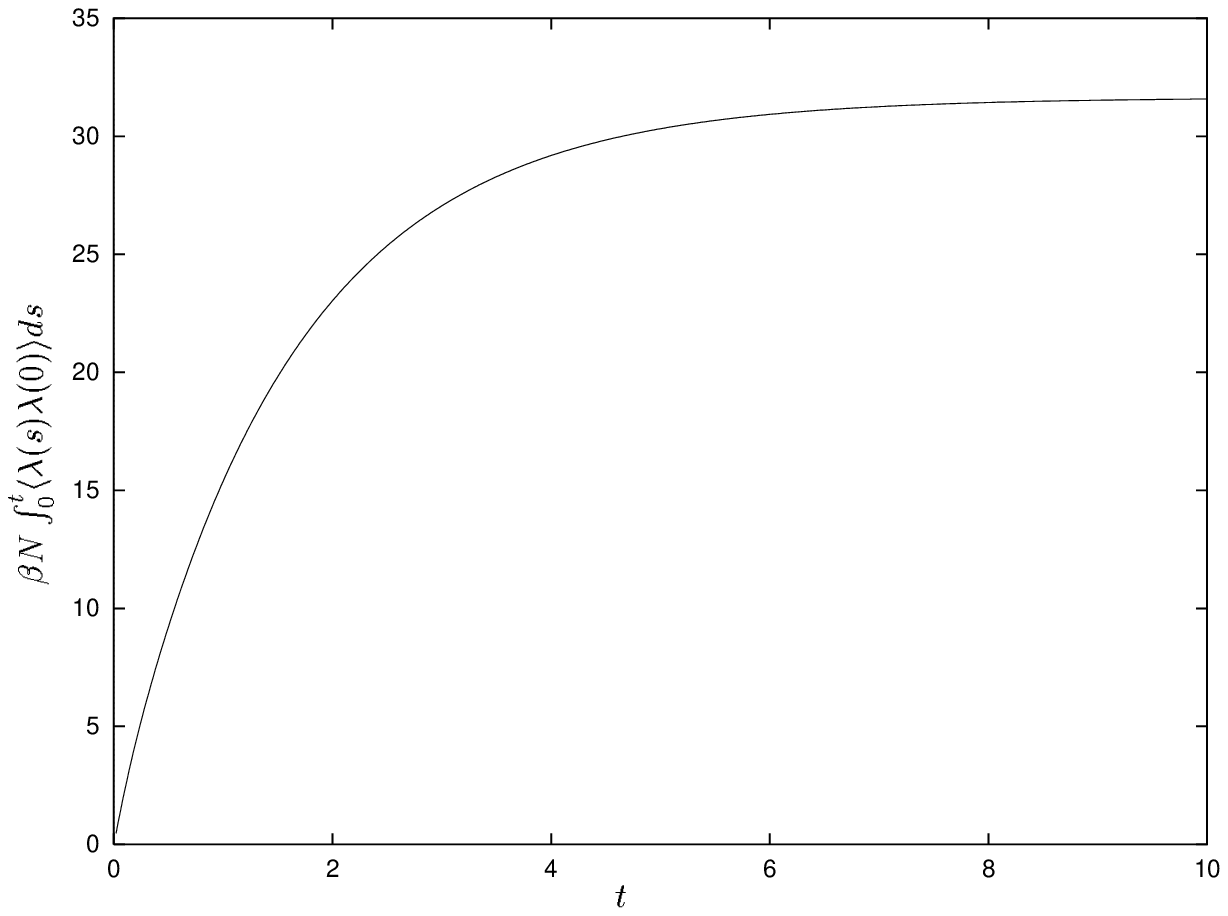}
\end{center}
\caption{Integral of the TTCF for the Linear Transport Coefficient}
\label{linear-fig}
\end{figure}

\begin{figure}
\begin{center}
\mbox{}\epsfbox{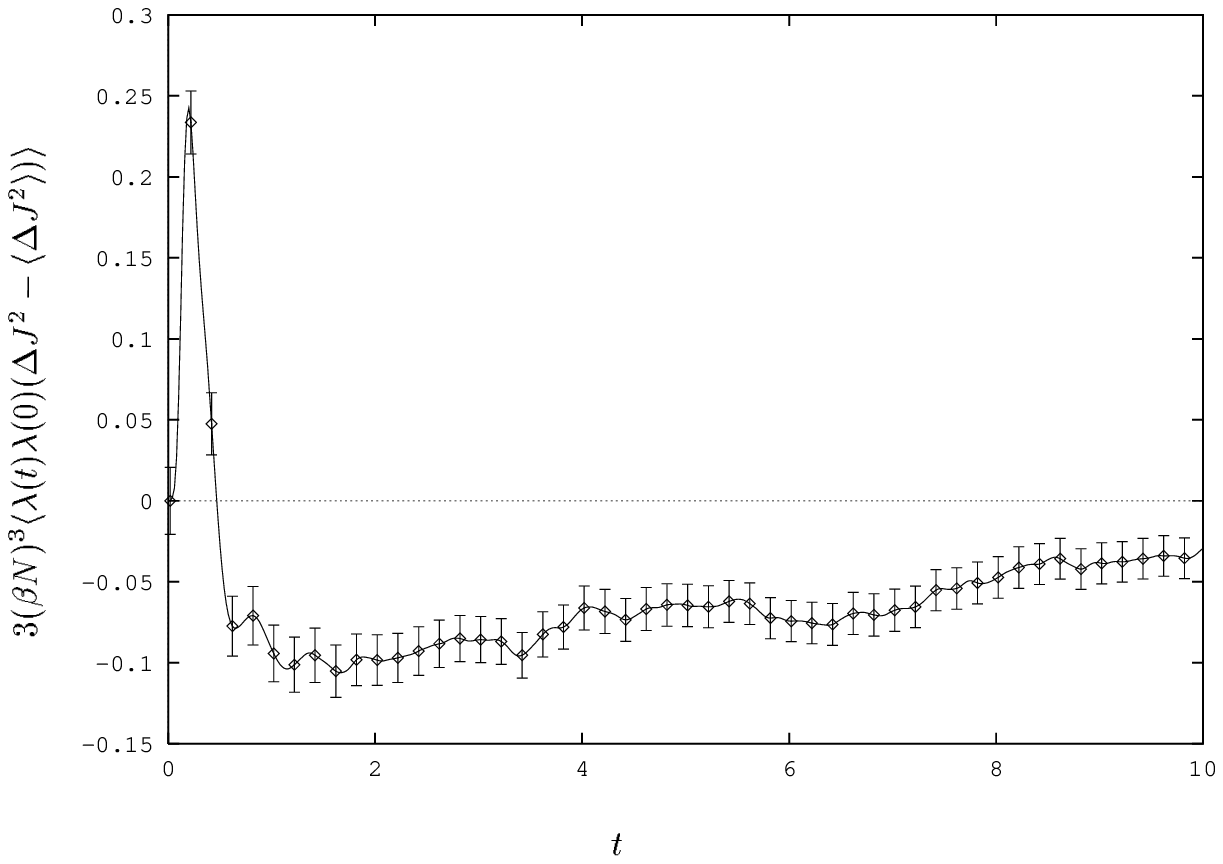}
\end{center}
\caption{Transient Time Correlation Function for the 32 particle
  system with $Q_\lambda=4.74$ at $1.32\times10^{11}$ timesteps}
\label{Fig1}
\end{figure}

\begin{figure}
\begin{center}
\mbox{}\epsfbox{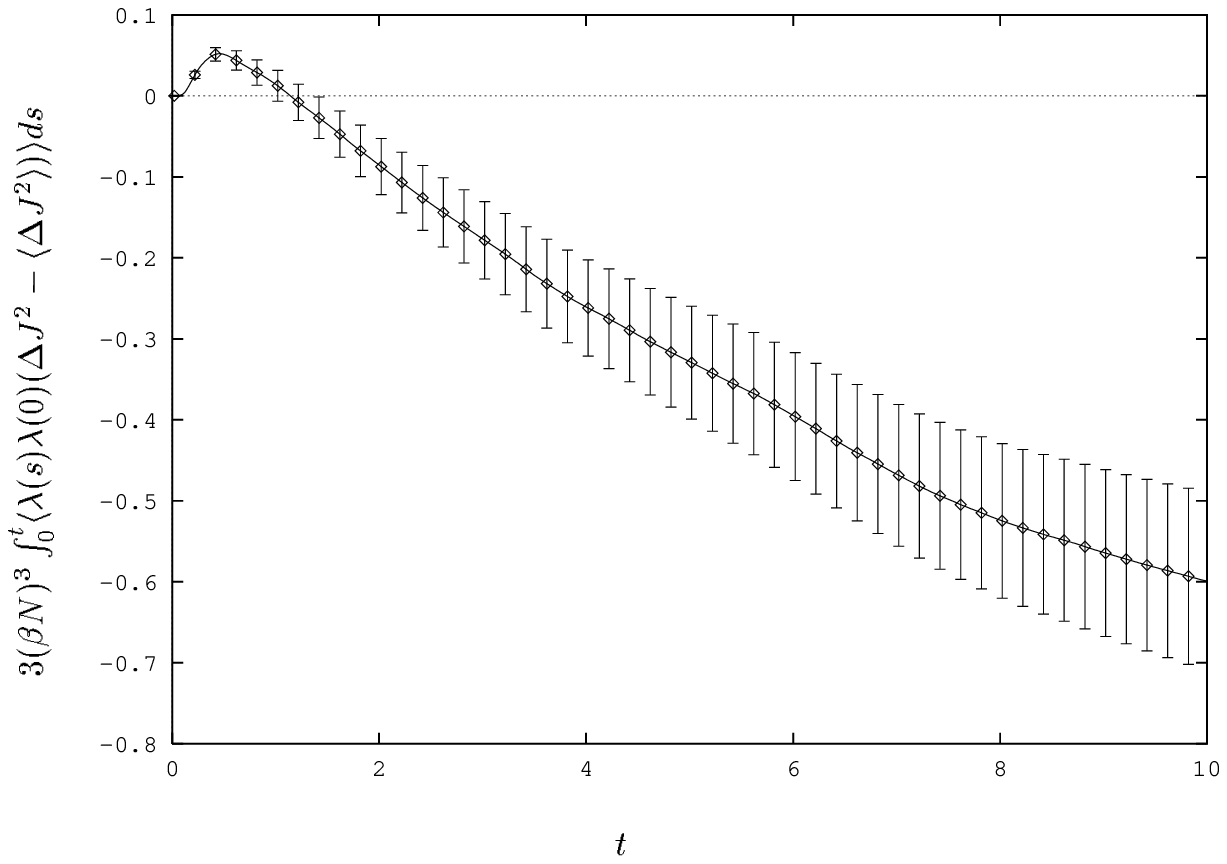}
\end{center}
\caption{Integral of TTCF for the 32 particle system with $Q_\lambda=4.74$ at $1.32\times10^{11}$ timesteps}
\label{Fig2}
\end{figure}

The TTCFs and their integrals are shown in figures
\ref{Fig1}--\ref{Fig6}. There is a considerable system size
dependence, indicating that the nonlinear Burnett coefficients diverge
in the thermodynamic limit, although the individual TTCFs remain
finite. It can be shown, using the lemma proved in the appendix of
\cite{Standish-Evans93}, that the inverse nonlinear Burnett
coefficients given by equation (\ref{B2}) should be intensive.  As well as
this, the 32 particle simulation shows strong evidence of a long time
tail (Fig \ref{Fig1} and \ref{Fig2}) when $Q_\lambda$ is increased
(softening the current-statting), leading to a divergence in the
integrals as $t\rightarrow\infty$. For comparison, the transient time
correlation function for the linear coefficient is shown in
Fig. \ref{linear-fig}, showing convergence within $t=5$.

\section{Conclusion}

This work presents strong numerical evidence in favour of infinite
nonlinear Burnett coefficients for soft spheres as is the case for
hard spheres. However, the Taylor series expansion of the constitutive
relation presented in \cite{Standish-Evans90} can also be derived for
$J_0\neq0$, which if the hard sphere model is anything to go by,
should be finite. These can be used to compute the constitutive
relation into the nonlinear region. However, it will probably be at
least another decade before these calculations become practical.

\section{Acknowledgements}

The author wishes to thank the New South Wales Centre for Parallel
Computing and the Fujitsu Parallel Computing Research Facilities for
use of their facilities for this work. He would also like to thank
Denis Evans for some helpful comments on the manuscript.



\begin{figure}
\begin{center}
\mbox{}\epsfbox{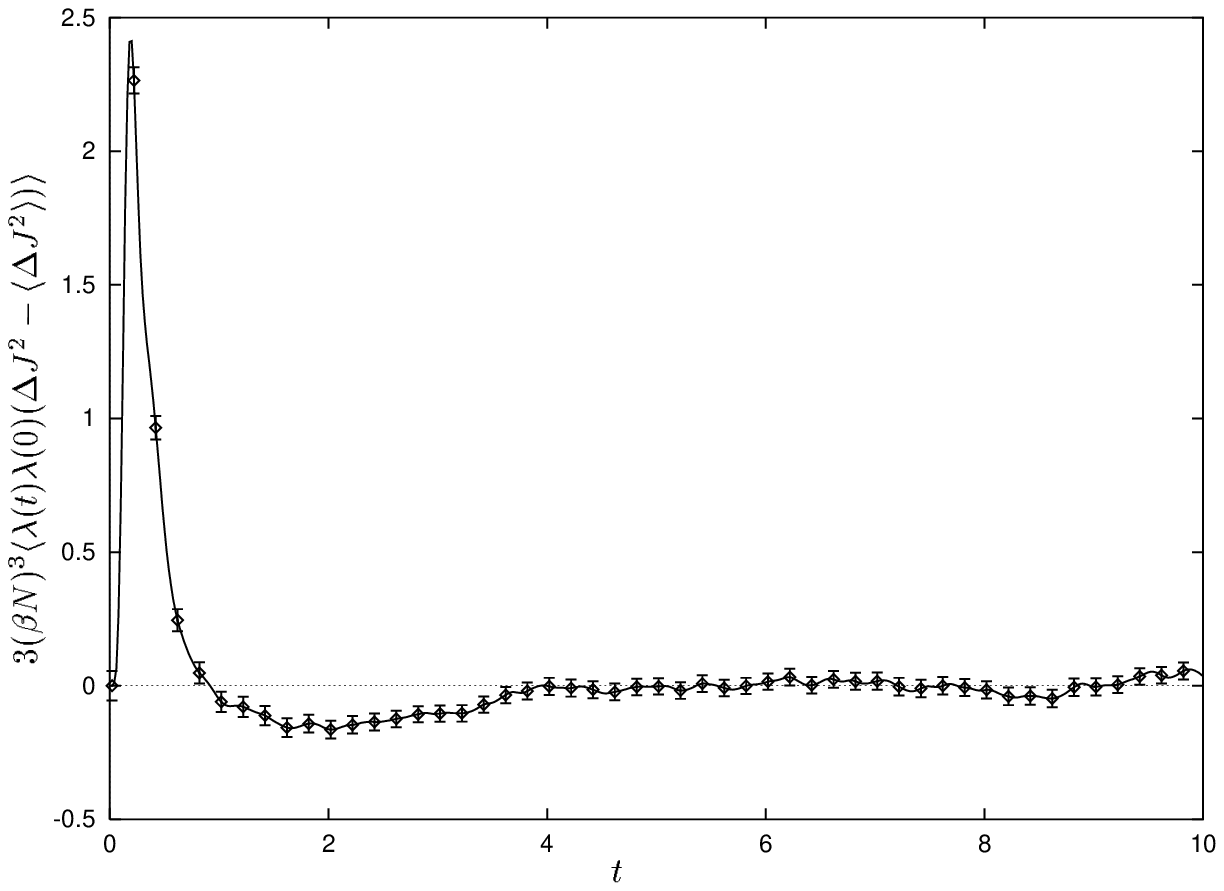}
\end{center}
\caption{Transient Time Correlation Function for the 32 particle
  system with $Q_\lambda=1.4$ at $2.2\times10^{11}$ timesteps}
\end{figure}

\begin{figure}
\begin{center}
\mbox{}\epsfbox{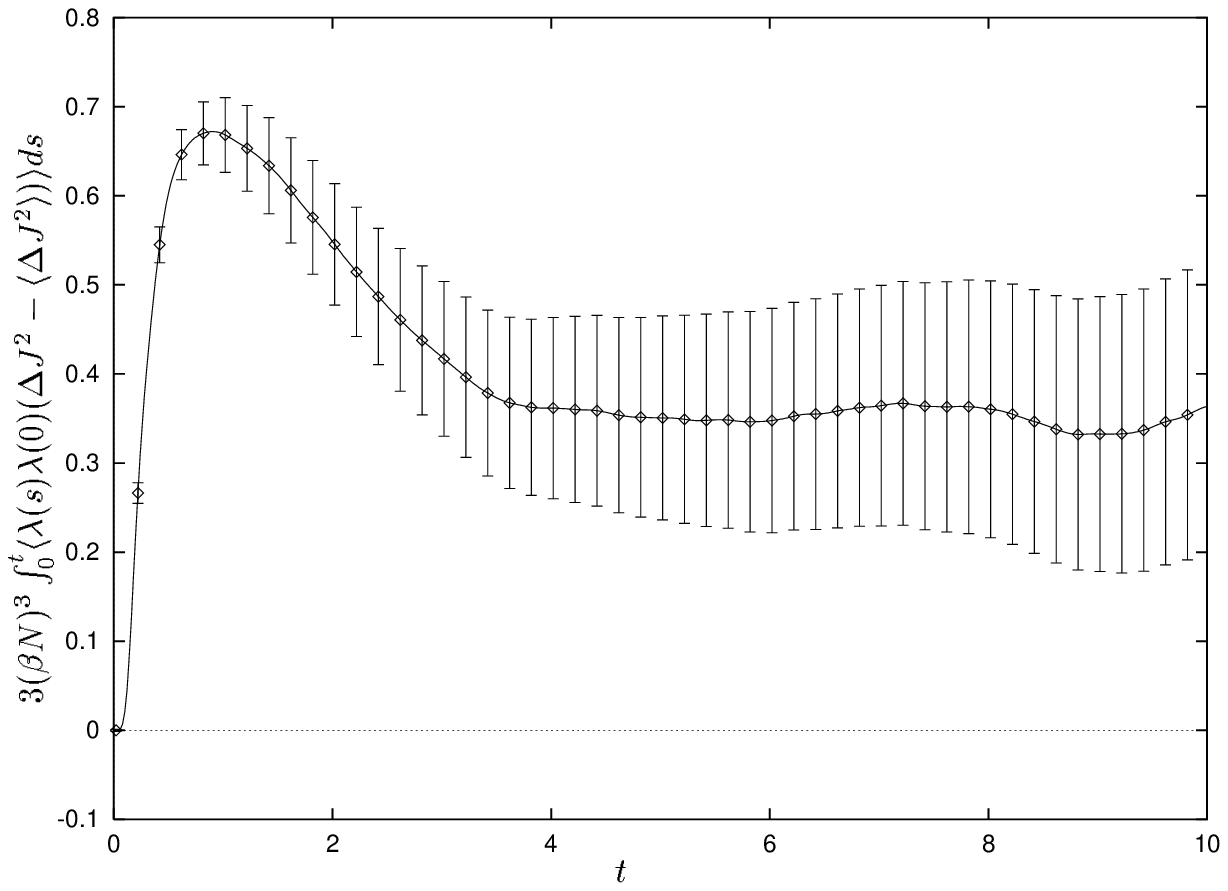}
\end{center}
\caption{Integral of TTCF for the 32 particle system with
  $Q_\lambda=1.4$ at $2.2\times10^{11}$ timesteps}
\end{figure}

\begin{figure}
\begin{center}
\mbox{}\epsfbox{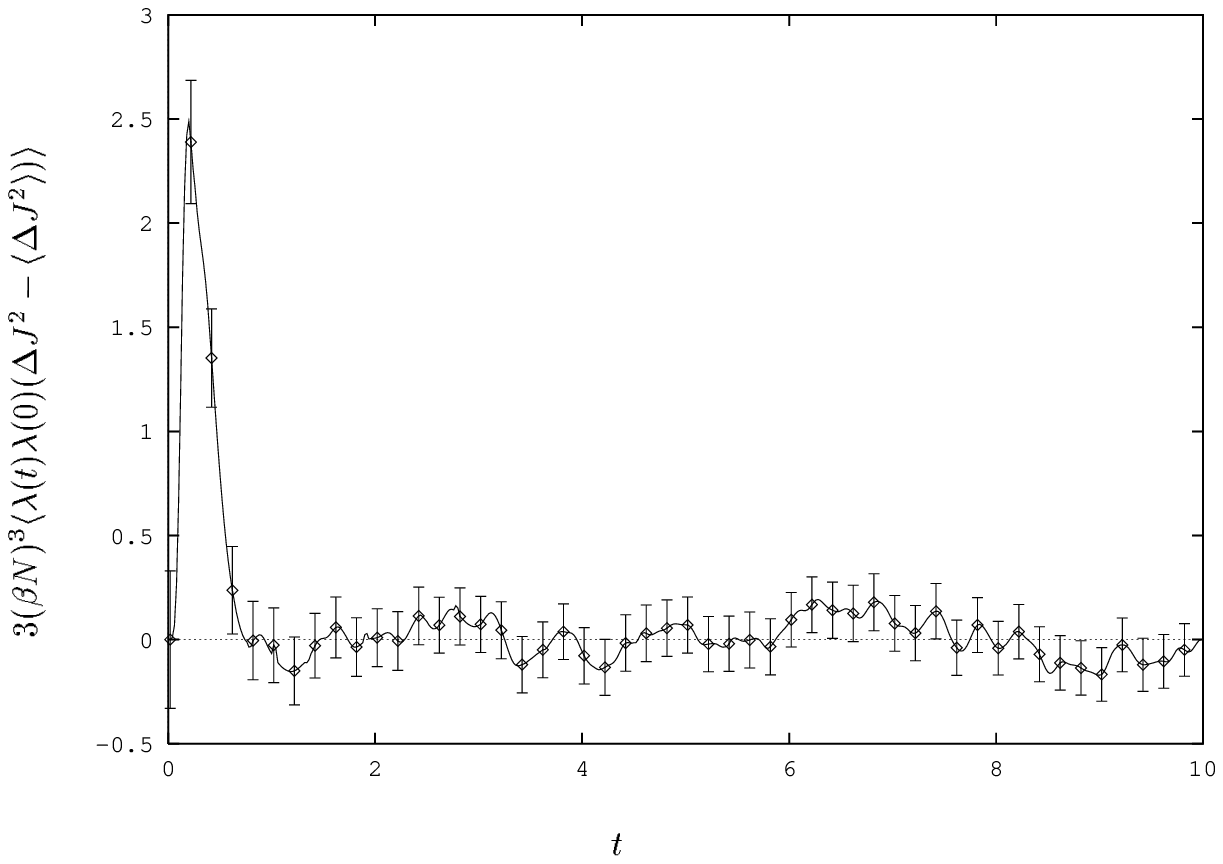}
\end{center}
\caption{Transient Time Correlation Function for the 108 particle
  system at $1.1\times10^{11}$ timesteps}
\end{figure}

\begin{figure}
\begin{center}
\mbox{}\epsfbox{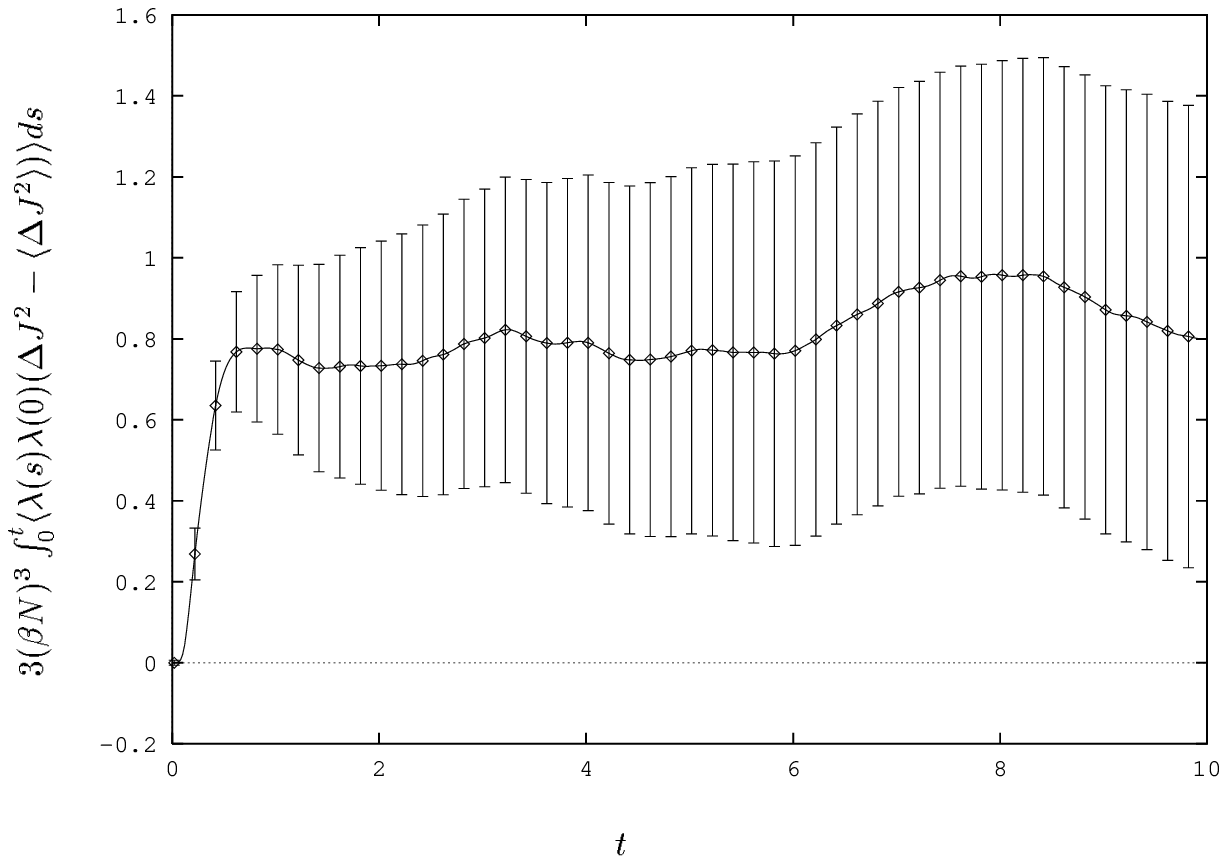}
\end{center}
\caption{Integral of TTCF for the 108 particle system at $1.1\times10^{11}$ timesteps}
\end{figure}

\begin{figure}
\begin{center}
\mbox{}\epsfbox{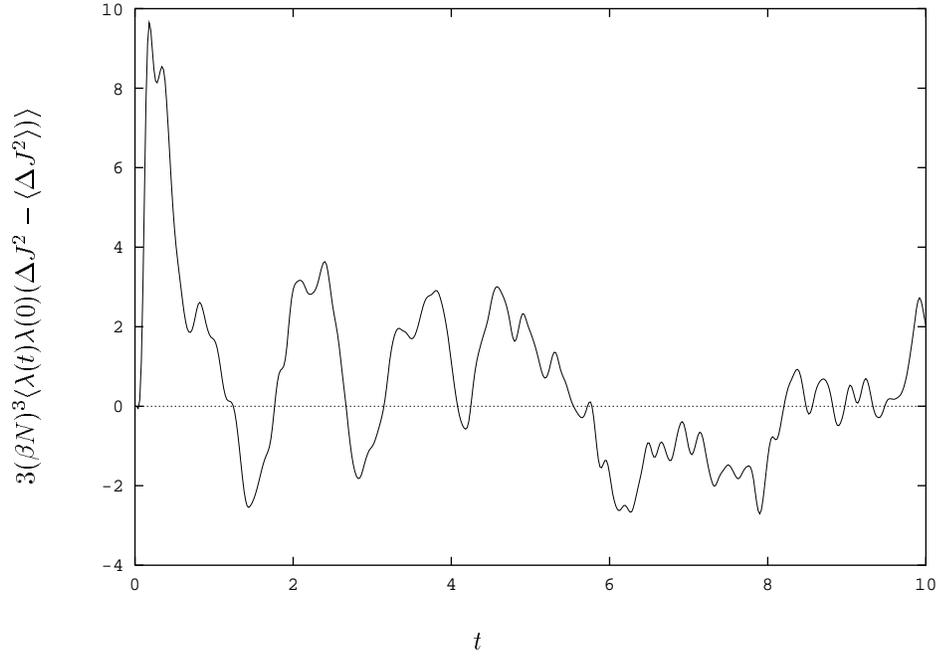}
\end{center}
\caption{Transient Time Correlation Function for the 256 particle
  system at $3\times10^{10}$ timesteps}
\end{figure}

\begin{figure}
\begin{center}
\mbox{}\epsfbox{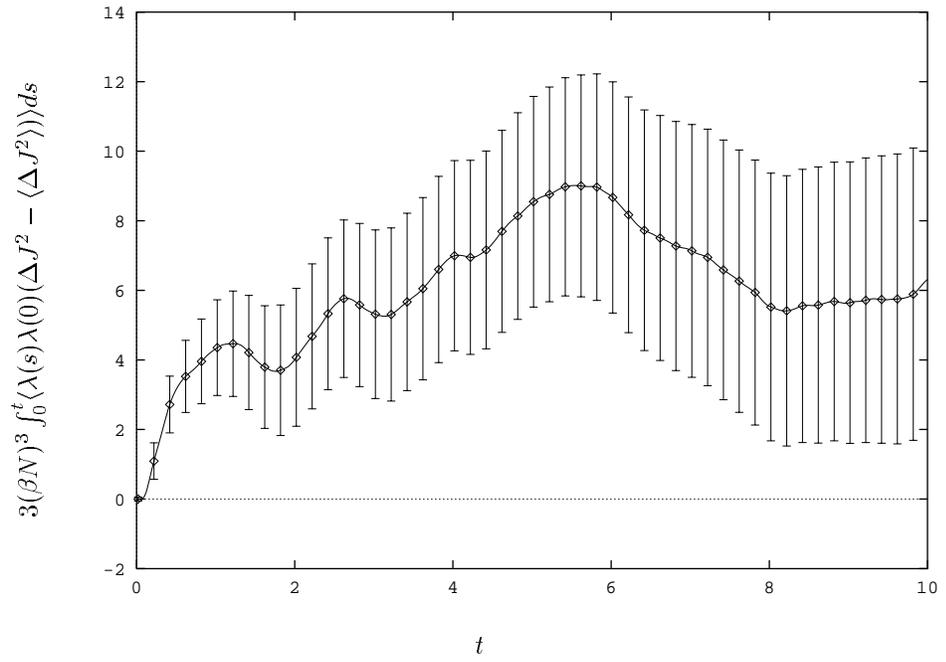}
\end{center}
\caption{Integral of TTCF for the 256 particle system at $3\times10^{10}$ timesteps}
\label{Fig6}
\end{figure}

\end{document}